\documentclass[a4paper, pre, 10pt, twocolumn, superscriptaddress]{revtex4}
\usepackage[english]{babel}
\usepackage[utf8]{inputenc}
\usepackage{graphicx,epsfig,placeins}
\usepackage{latexsym}
\usepackage{hyperref}
\usepackage{amsfonts,amssymb,bm,amsmath}
\usepackage{textcomp}
\usepackage{epstopdf}
\usepackage{color}
\usepackage{appendix}
\usepackage{stackengine}

\newcommand{\ff}[2]{\frac{#1}{#2}}

\def\about{\raise.17ex\hbox{$\scriptstyle\mathtt{\sim}$}}
\def\ket|#1>{| #1 \rangle}
\def\bra<#1|{\langle #1 |}
\def\<{\langle}
\def\>{\rangle}
\def\({\left(}
\def\){\right)}
\def\[{\left[}
\def\]{\right]}
\def\{{\left\lbrace}
\def\}{\right\rbrace}
\def\pl{\partial}

\def\R{{\mathbb R}}
\def\e{{\mathrm e}}

\begin{document}

\title{A null model for Dunbar's circles}

\author{Manuel Jiménez-Martín}
\affiliation{Dpto. F\'isica Fundamental, Universidad Nacional de
  Educación a Distancia, Madrid (Spain).}

\author{Silvia N. Santalla}
\affiliation{Dpto. Física y Grupo Interdisciplinar de Sistemas
  Complejos (GISC), Universidad Carlos III de Madrid, Leganés (Spain).}

\author{Javier Rodríguez-Laguna}
\affiliation{Dpto. F\'isica Fundamental, Universidad Nacional de
  Educación a Distancia, Madrid (Spain).}

\author{Elka Korutcheva}
\affiliation{Dpto. F\'isica Fundamental, Universidad Nacional de
  Educación a Distancia, Madrid (Spain).}
\affiliation{Dep. Theoretical Physics, G. Nadjakov Institute of Solid
  State Physics, Bulgarian Academy of Sciences, 72 Tzarigradsko
  Shaussee Blvd. 1784 Sofia (Bulgaria).}

\begin{abstract}
An individual's social group may be represented by their ego-network,
formed by the links between the individual and their
acquaintances. Ego-networks present an internal structure of
increasingly large nested layers (or circles) of decreasing
relationship intensity, whose size exhibits a precise scaling
ratio. Starting from the notion of limited social bandwidth, and
assuming fixed costs for the links in each layer, we propose a null
model built on a grand-canonical ensemble that generates the observed
hierarchical social structure. The observed internal structure of
ego-networks becomes a natural outcome to expect when we assume the
existence of layers demanding different amounts of resources. In the
thermodynamic limit, reached when the number of ego-network copies is
large, the specific layer degrees follow a Poisson distribution. We
also find that, under certain conditions, equispaced layer costs are
necessary to obtain a constant group size scaling. Our model presents
interesting analogies to a Bose-Einstein gas, that we briefly
discuss. Finally, we fit and compare the model with an empirical
social network.
\end{abstract}

\date{November 30, 2019}
\maketitle

\section{Introduction}

The computational capacity to store and manage an ever-changing social
network is thought to depend roughly on neocortical size, which
evolved driven by the need of managing increasingly large social
groups \cite{dunbar1992}. This is the statement of the far reaching
\emph{social brain hypothesis} \cite{dunbar1998}, which links brain
volume in humans, primates and other mammals with the size of their
social groups.
For humans, Dunbar's number constitutes an upper limit of $\about150$
for the social group size. That is, the total number of active
relationships that we can maintain at any given time; a cognitive
constraint that seems to operate also in virtual environments, such as
Twitter \cite{GoncalvesVespignaniPONE2011}.
Certainly, monitoring and handling social ties comes at a cost, since
it takes time to cultivate these relationships
\cite{MiritelloMoroSR2013}, and the process is limited by cognitive
constraints, such as memory and mentalising skills
\cite{StillerDunbarSN2007}.
Current sociological studies rely on large digital datasets in order
to build weighted networks of human relationships, where link weights
encoding actor-to-actor interaction frequency are used as a proxy for
emotional closeness \cite{RobertsDunbarEHB2011, ArnaboldiCC2013}.  
In this framework, an individual's social group is equivalent to its
set of neighbors, which is often called its {\em ego-network}.

\begin{figure}
  \includegraphics[width=7cm]{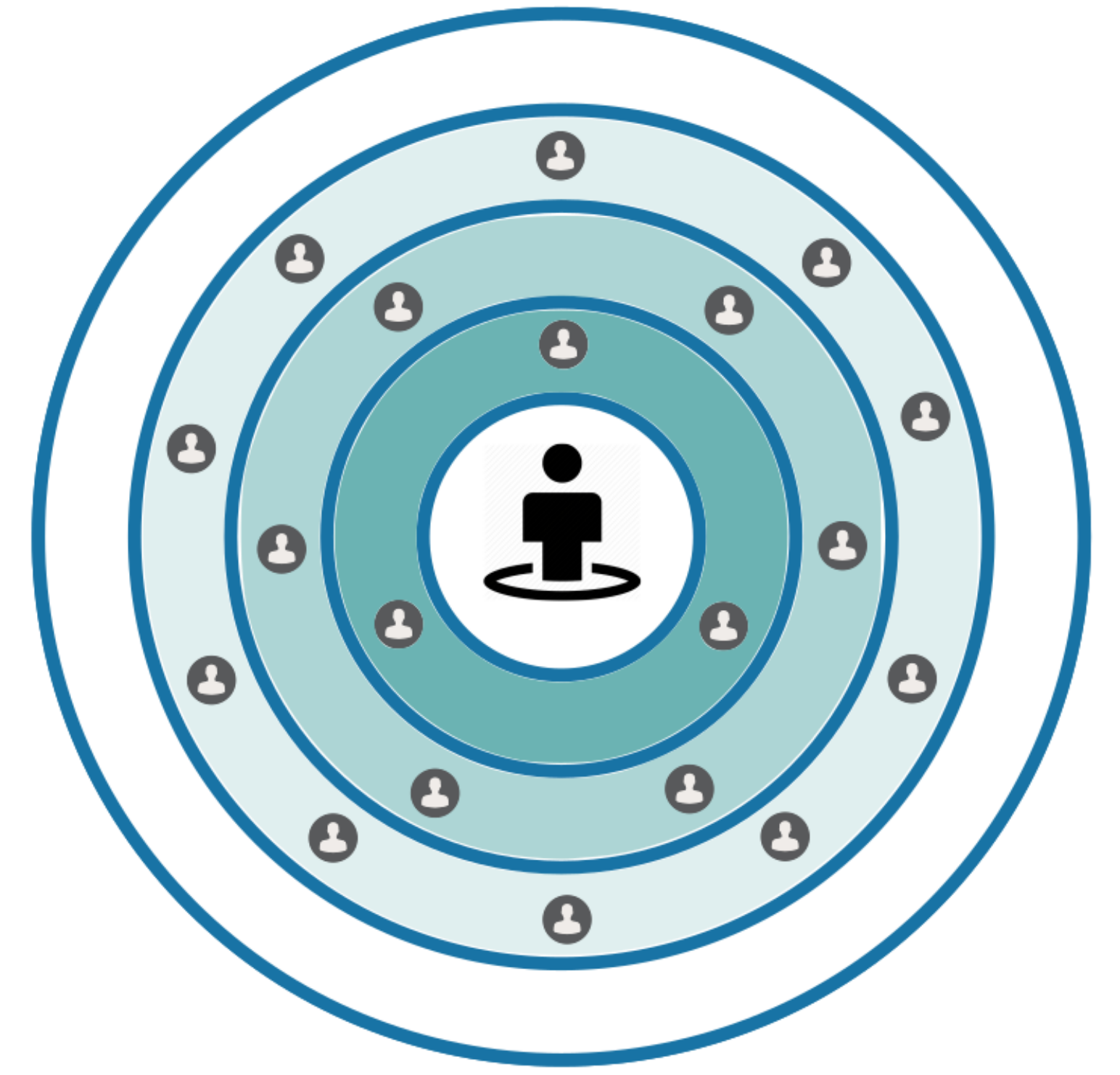}
  \caption{Illustrating the ego network and the Dunbar circles.}
  \label{fig:dunbar}
\end{figure}

Ego-networks are internally highly structured and their links can be
sorted by their weights \cite{SaramakiDunbarPNAS2014}. Moreover, links
can be clustered into groups of increasing number of links and
decreasing emotional closeness \cite{ZhouDunbar2005}.
These layers form a nested hierarchy, where the cumulative sizes of
consecutive groups follow a preferred scaling ratio of approximately
$3$, forming a sequence of typical group sizes of $\about5$,
$\about15$, $\about50$ and $\about150$, which are sometimes called
{\em Dunbar's circles}, as illustrated in Fig. \ref{fig:dunbar}.
A smaller inner layer of size $\about1.5$
\cite{DunbarPasarellaSN2015}, and two larger groups of sizes
$\about500$ and $\about1500$ \cite{DunbarBBS1993}, have also been
reported. In each case, the scaling relationship between consecutive
groups holds.
This hierarchical structure appears to be a fundamental organizational
principle of human groups, and has been confirmed in online games
\cite{FuchsThurner2014}, online social networks \cite{ArnaboldiFB2012,
  ArnaboldiTW2013} and telephone call detail records
\cite{MacCarronDunbarSN2016}.

Null models constitute a fundamental tool in statistical analysis
\cite{Null_Models}. A null model is an ensemble defined by a few
simple constraints, such that its random samples reproduce some
properties of naturally occurring objects. The purpose of null models
is to determine which statistical properties of our natural objects
are a simple consequence of the aforementioned constraints, and which
ones are not. Several null models suitable for weighted social
networks have been proposed recently \cite{SagarraGuileraPRE2013,
  SquartiniGarlaschelliSR2013,SagarraGuileraPRE2015}.

In this paper, we propose a grand-canonical null model that reproduces
qualitatively the hierarchical structure of ego-networks, and is able
to fit experimental data successfully.
Our constraint is to associate a constant cost to the social ties
within each layer, and to postulate an abstract social capital
\cite{antonioni2014reds}, or resource, that is spent in placing the
links into the different layers.
Then, by fixing the actors' average total degree and resource, the
hierarchical structure emerges spontaneously.
This ensemble is an unbiased null model for ego-networks which offers
a parsimonious explanation for the nested structure. It can also be
used to generate synthetic data with the desired layer scaling, as
well as for hypothesis testing against more complex models. Indeed, an
earlier version of this work \cite{older} was relevant for researchers
analyzing ego-networks from real social data \cite{Anxo}.

Our null model presents a formal analogy with Bose-Einstein
statistics, with the layers playing the role of energy levels and the
acquaintancies playing the role of bosonic particles
\cite{KARDAR,Pitaevski,Grossmann}. Indeed, Bose-Einstein condensation
in complex systems has been reported earlier in the context of
evolving networks despite their non-equilibrium and irreversible
nature \cite{Bianconi}. Within the evolving model and assigning an
energy to each node, determined by a certain fitness parameter, as
well as by assuming that each node increases its connectivity
following a power-law, the authors were able to show that the system
can be mapped in the thermodynamic limit to an equilibrium Bose gas.

The ensemble probability distribution, as well as its thermodynamic
limit, is presented in section \ref{s:grand_canonical_ensemble}. In
section \ref{s:hierarchical} we examine the hierarchical structure and
prove that, in our setting, equispaced costs are a condition for a
constant group size scaling in the outer layers. We fit and compare
the model to an empirical dataset in section
\ref{ss:RS_dunbar_structure}. Finally, section \ref{s:discussion} is
devoted to the discussion of the results.


\section{Grand-canonical ensemble for Dunbar's circles}
\label{s:grand_canonical_ensemble}

Let us assume the existence of a social resource, $s$, that an
individual can employ to establish ties or relationships of different
emotional intensity with their $k$ acquaintances.
We consider $r=1,\dots,R$ different relationship layers with
respective costs, $s_r\in\R$, with $0 \leq s_r \leq s$ and sorted in
order of  decreasing emotional intensity, such that $s_r > s_{r+1}$.   
We define the \emph{layer-degree}, $k_r$, as the number of ties of
cost $s_r$.  
In sum, a given individual identified by index $j$ will have a total
degree and social resource verifying 
\begin{subequations}\label{eq:micro-canonical_constraints}
	\begin{alignat}{2}
	 k(j) &= \sum_r k_r(j), \\ 
	 s(j) &= \sum_r s_r k_r(j) .
	\end{alignat}
\end{subequations}
Dunbar's circles are inclusive groupings of decreasing emotional
closeness \cite{ZhouDunbar2005}, hence the {\em group at level} $r$
includes all layers of cost higher or equal to $s_r$. In our
setting, the variables corresponding to the size of Dunbar's circles
are the cumulative \emph{group-sizes}, defined as
$n_r=\sum_{l=1}^r k_l$.

Thus, an individual's ego-network in our setting is completely
described by the configuration variables $k_r$.
This system can take any state, $\{k_r\}_{r=1,\dots,R}$ that verifies
the constraints Eqs. \eqref{eq:micro-canonical_constraints}.  
The problem is equivalent to that of distributing $k$ particles among
$R$ energy levels $\{s_r\}$ in a quantum bosonic micro-canonical
ensemble \cite{KARDAR,Pitaevski}.
Without making any assumption \emph{a priori} about the probabilities
of any given configuration, we could study the average layer structure
in a micro-canonical ensemble. The micro-canonical ensemble assigns an
homogeneous probability distribution on the configuration space, hence
the problem would be solved if we could compute the total number of
allowed configurations, i.e. the partition function. 
However, as in classical statistical mechanics, it is simpler to
formulate instead a generalized or grand-canonical ensemble,
consisting on a large number of identical copies of the system, for
which the constraints Eqs. \eqref{eq:micro-canonical_constraints} are
verified only on average, formally leading to Bose-Einstein
probability distribution functions.

Consider a group of $N$ individuals or egos, each of which having
different degree, $k(j)$ and social resource $s(j)$; with given averages   
\begin{subequations}\label{eq:grand_canonical_constraints}
\begin{alignat}{2}
 \<k\> &= \sum_{j=1}^N k(j) P(\{k_r(j)\}), \\
 \<s\> &= \sum_{j=1}^N s(j) P(\{k_r(j)\}). 
\end{alignat}
\end{subequations}

The least biased distribution $P(\{k_r\})$ verifying the constraints
Eqs. \eqref{eq:grand_canonical_constraints} can be calculated
following a maximum entropy principle \cite{JAYNES}.  The distribution
entropy is $S=-\sum_{\{k_r\}}P(\{k_r\})\ln P(\{k_r\})$, where the sum
runs over all the allowed configurations.  Maximizing $S$ subject to
the constraints Eqs. \eqref{eq:grand_canonical_constraints} plus
normalization, we obtain a Gibbs distribution 
\begin{equation}
	P(\{k_r\}) = \frac{1}{Z} D(\{k_r\})\e^{H(\{k_r\})},
\end{equation}
where $D(\{k_r\}) = \binom{N}{k} k!/\prod_r k_r!$ is the degeneracy of
the configuration $\{k_r\}$, which counts all possible ways of
selecting $k=\sum_r k_r$ links out of $N$ actors, and all ways of
assigning $k$ distinguishable links into layers of degree
$k_r$. Finally, $Z$ and $H$ are the partition function and cost
function, respectively. 
\begin{align}
  \centering
  Z &= \sum_{\{k_r\}} D(\{k_r\}) \e^{H(\{k_r\})}, \\
  H(\{k_r\}) &= \lambda k + \mu s  + \textstyle\sum_r h_r k_r\displaystyle.
\end{align}
Here $\lambda$ and $\mu$ are the Lagrange multipliers and will act as
fitting parameters in order to enforce the constraints
Eqs. \eqref{eq:grand_canonical_constraints}. We have also included the
auxiliary fields $h_r$ for convenience.

The maximum entropy method has been applied to formulate a large
number of complex network models with prescribed features, known
generally as exponential random graphs
\cite{ParkNewmanPRE2004,GarlaschelliLoffredoPRE2008,BianconiPRE2009,
  SagarraGuileraPRE2013,SagarraGuileraPRE2015}. Our approach here
seeks instead a probability distribution for ego-network
configurations $\{k_r\}$. All the information of our system, including
the cumulants of the layer degrees and group sizes, as well as their
marginal distributions, are recovered from the partition function,
which can be calculated analytically:
\begin{subequations}\label{eq:_partition_function}
\begin{alignat}{3}
 &Z = \sum_{\{k_r\}} \binom{N}{k}\frac{k!}{\prod_r k_r!} \e^{\sum_r(\lambda + \mu s_r + h_r)k_r} \\
 &= \sum_{k=0}^N \binom{N}{k}  \sum_{\{k_r|k\}} k! \prod_r\frac{\e^{(\lambda + \mu s_r + h_r)k_r}}{k_r!} \\
 &= \sum_{k=0}^N \binom{N}{k}  \( \sum_r \e^{\lambda + \mu s_r + h_r} \)^k 
  = \( 1 + \sum_r \e^{\lambda + \mu s_r + h_r} \)^N. &
\end{alignat}
\end{subequations}
Here the symbol $\sum_{\{k_r|k\}}$ on the second line denotes sums
over configurations $\{k_r\}$ with total degree $k$, and we have used
the multinomial and binomial sums on the second and third lines,
respectively. 

The cumulants of a single layer degree $k_r$ can be computed by taking
derivatives with respect to the respective auxiliary field $h_r$. For
instance, the average layer degree, variance, as well as the
correlations between different layers are given by 
\begin{eqnarray}
\<k_r\> =& \pl_{h_r} \ln Z \mid_{h=0} &= \frac{N xy^{s_r}}{1+\sum_l xy^{s_l}}, \label{eq:canonical_mean_layer_degree}\\ 	
\sigma^2_{k_r} =& \pl^2_{h_r} \ln Z \mid_{h=0} &= \frac{1+\sum_{l\neq r}xy^{s_l}}{1+\sum_l xy^{s_l}} \<k_r\>, \label{eq:canonical_layer_degree_variance}\\
\sigma_{k_r, k_l} =& \pl^2_{h_r, h_l} \ln Z \mid_{h=0} &= -\frac{\<k_r\>\<k_l\>}{N}, \label{eq:canonical_layer_degree_corr} 
\end{eqnarray}
where we have defined $x=\e^{\lambda}$ and $y=\e^{\mu}$, and $h=0$
implies that all $h_r$ are set to zero.
We will prove later on that the layer degree marginal distributions
become uncorrelated Poisson distributions in the thermodynamic limit
$N\to\infty$.  
Let us first write, however, the saddle point equations, used to fix
the $k$ and $s$ ensemble averages.
\begin{align}
\<k\>  &= \pl_{\lambda} \ln Z \mid_{h=0}  = N\frac{\sum_r xy^{s_r}}{1+\sum_r xy^{s_r}}, \label{eq:saddle_k_canonical}\\
\<s\>  &= \pl_{\mu}  \ln Z \mid_{h=0}   = N\frac{\sum_r s_r xy^{s_r}}{1+\sum_r xy^{s_r}}. \label{eq:saddle_s_canonical}
\end{align}
Notice that the average degree and resource verify $\<k\>=\sum_r
\<k_r\>$ and $\<s\>=\sum_r s_r \<k_r\>$, respectively. In our maximum
entropy setting, Eqs. \eqref{eq:saddle_k_canonical} and
\eqref{eq:saddle_s_canonical} are solved for $x$ and $y$ in order to
obtain the parameter values that fix the desired $\<k\>$ and $\<s\>$.

Further derivatives recover the subsequent $k$ and $s$ cumulants. For
instance, the variances are 
\begin{align}
\sigma_k^2  &= \pl_{\lambda}^2 \ln Z\mid_{h=0}  = \frac{\<k\>}{1+\sum_r xy^{s_r}}, \\
\sigma_s^2  &= \pl_{\mu}^2  \ln Z\mid_{h=0}    =  \sum_r s_r^2 \<k_r\> -\frac{\<s\>^2}{N}. 
\end{align}
Finally, we can write the configuration probability function, which
may be sampled with Monte Carlo methods:
\begin{equation}
  P(\{k_r\})= \binom{N}{k}\frac{k!}{\prod_r k_r!}\cdot\frac{\prod_r
    (xy^{s_r})^{k_r}}{\(1+\sum_r xy^{s_r}\)^N}
  \label{eq:canonical_prob}.
\end{equation}
%


\subsection*{Thermodynamic limit}
\label{s:ensemble_equiv}

Let us study the thermodynamic limit for the grand-canonical ensemble,
that is when the number of ego-networks $N\to\infty$ while keeping
$\<k\>$ and $\<s\>$ constant. From Eq. \eqref{eq:saddle_k_canonical}
we can write $\sum_r xy^{s_r} = \<k\>/(N-\<k\>)$. This in turn, allows
us to rewrite the partition function as
\begin{equation}
Z=\(1+\frac{\<k\>/N}{1-\<k\>/N}\)^N\; \xrightarrow[N\to\infty] \; \e^{\<k\>}=\prod_r \e^{\<k_r\>}.
\end{equation}
The expected layer degrees can be expressed as
$\<k_r\>=xy^{s_r}(N-\<k\>) \cong Nxy^{s_r} $, for which it is needed
that $xy^{s_r}\to0$ as $N\to\infty$. Then, the configuration
probability distribution Eq. \eqref{eq:canonical_prob}, reduces to 
\begin{equation}
  P(\{k_r\}) = \frac{N^{-k} N!}{(N-k)!} \prod_r \frac{\<k_r\>^{k_r}}{k_r!}\e^{-\<k_r\>} \; \xrightarrow[N\to\infty] \; \prod_r p_r(k_r),
\end{equation}
where $p_r(k_r)$ are the layer degree marginal distributions, and the
prefactor tends to $1$ as $N\to\infty$, for finite $k$. Indeed, using
Stirling's approximation and the exponential limit
\begin{equation}
\frac{N^{-k}N!}{(N-k)!} \cong \e^{-k}\(\frac{N}{N-k}\)^{N-k} \xrightarrow[N\to\infty]\, 1.
\end{equation}
Thus, the resulting layer degree marginal distributions are Poisson
distributions.
\begin{equation}
p_r(k_r) =P(k_r|N,x,y)=\frac{\e^{N xy^{s_r}}}{k_r!}(N xy^{s_r})^{k_r}.
\end{equation}
Noticing that $\sum_r xy^{s_r}\to0$, we can see that
Eqs. \eqref{eq:canonical_mean_layer_degree},
\eqref{eq:canonical_layer_degree_variance} and
\eqref{eq:canonical_layer_degree_corr} reduce to
\begin{eqnarray}
\<k_r\> &=& N x y^{s_r}, \label{eq:micro-canonical_mean_layer_degree} \\
\sigma^2_{k_r} &=& \<k_r\>, \label{eq:micro-canonical_layer_degree_variance}\\
\sigma_{k_r, k_l} &=& 0. \label{eq:micro-canonical_layer_degree_corr}
\end{eqnarray}
Moreover, the saddle point equations,
Eqs. \eqref{eq:saddle_k_canonical} and \eqref{eq:saddle_s_canonical},
become 
\begin{align}
 k &= \sum_r     Nx y^{s_r}, \label{eq:saddle_k}\\
 s &= \sum_r      s_r\,Nx y^{s_r}. \label{eq:saddle_s}
\end{align}

Thus, as we anticipated, the ego-network grand-canonical ensemble
generates an uncorrelated layer structure in the thermodynamic
limit. Hence, sampling corresponds to drawing independent random
Poisson variables from the layer degree marginal distributions.


\section{Hierarchical structure}
\label{s:hierarchical}


Before proceeding to study the constant group size scaling condition,
let us first discuss the meaning of the ensemble parameters, $x$ and
$y$.
Both $x$ and $y$ are positive, since they are defined as exponentials
of the real Lagrange multipliers $\lambda$ and $\mu$.  
We can identify $y^{s_r}$ as the relative weight of each layer when
writing the average link weight 
\begin{equation}
\bar s = \ff{s}{k} = \frac{\sum_r s_r y^{s_r}}{\sum_r y^{s_r}}.	
\end{equation}
Moreover, the parameter $y$ relates the expected layer-degree scaling
with the difference of the link costs: 
\begin{equation}\label{eq:layer_degree_scaling}
\frac{\<k_{r+1}\>}{\<k_r\>}=y^{s_{r+1}-s_r}.	
\end{equation}
On the other hand, $x$ acts a volumetric parameter which fixes the
total degree through the constant product $Nx=k/\sum_r y^{s_r}$. 

From equation \eqref{eq:layer_degree_scaling}, the hierarchical
structure is made apparent, that is $\<k_r\> < \<k_{r+1}\>$, as long
as $y<1$.  

The value of $y$ depends on $\bar s$. It can be shown that $y<1$ if
$\bar s<\sum_r s_r/R$ and $y>1$ if $\bar s>\sum_r s_r/R$.  That is, if
the average link weight is smaller than the average layer cost, outer
layers will have have increasing number of links.  The case $y>1$
defines an inverse regime that can be applied to ego-networks where a
fixed social capital $s$ must be distributed among relatively few
links $k$, and most ties belong to the inner layers.  The inverse
regime has been succesfully applied recently to model personal
networks of inmigrants \cite{Anxo}. However, on this paper we will
focus on the $y<1$ regime, which models correctly layers of unbounded
increasing cost.

Let us check that this is indeed the case. Notice that the partition
function $Z$ measures the number of allowed configurations and that
the layer costs $s_r$ are arbitrary positive numbers.
Consider an increase in the cost of one of the layers, $s_r$, while
keeping the imposed average values, $\<k\>$ and $\<s\>$ constant.  For
large enough $s_r$ and fixed $\<s\>$, an overwhelming majority of
ego-networks will not have any links placed in layer $s_r$, and the
number of allowed configurations must decrease. In the limit
$s_r\to\infty$, a well behaved partition function demands $\mu<0$ and,
consequently, $y\in[0,1]$: the layer degrees $\<k_r\>$ are
monotonically decreasing with the layer cost $s_r$.

In sum, in a maximum entropy setting corresponding to the least
unbiased guess about the ego-network configurations, a hierarchical
structure arises naturally from the constraints
Eqs. \eqref{eq:grand_canonical_constraints}. Next, let us consider the
condition of a constant group size scaling, as it is observed on
empirical ego-networks.


\subsection*{Constant group size scaling condition}
\label{s:group_size_scaling}

In human social groups, a constant scaling is found on average between
the cumulative sizes of consecutive layers.
In the grand-canonical ensemble, this is expressed by the expected
group-size scaling, $\<{n_r}/{n_{r+1}}\>$. This quantity is the
quotient of two functions of the configuration variables which can be
approximated in terms of $\<k_r\>$, $\<k_r^2\>$ and $\<k_r k_s\>$, as
explained in Appendix \ref{s:appendix_quotient_cumulants}. The
ensemble average is given by
\begin{equation}
\left< \ff{n_r}{n_{r+1}} \right> = \ff{\<n_r\>}{\<n_{r+1}\>}(1+\epsilon_{r+1}), 
\end{equation} 
where $\epsilon_{r+1}$ is a second order correction term, which can be
expressed as 
\begin{equation}
 \epsilon_{r+1}  
 =  \frac{\<n_r\>\<k_{r+1}^2\>-\<k_{r+1}\>\<n_r^2\>}{\<n_r\>\<n_{r+1}\>^2}
 =  \frac{\<k_{r+1}\>^2-\<k_{r+1}\>\<n_r\>}{\<n_{r+1}\>^2}.
 \end{equation} 
The rightmost expression was obtained by using the identity for
Poisson variables: $\<k_r^2\>=\<k_r\>+\<k_r\>^2$.

Let us now consider the scaling of two consecutive group pairings
$\<n_{r+1}/n_r\>$ and $\<n_r/n_{r-1}\>$. We will consider the
implications of having a constant group-size scaling, such as observed
in empirical relationship networks. Imposing
$\<n_{r+1}/n_r\>=\<n_r/n_{r-1}\>$ we get
\begin{equation}
  \<n_r^2\>=\<n_{r+1}\>\<n_{r-1}\>R_r.
\end{equation}
The correction factor, $R_r=(1+\epsilon_r)/(1+\epsilon_{r+1})$, tends
to 1 provided that $\<n_r\> \gg 1$. Indeed, this is a good
approximation for the outer layers, as shown in the inset of
Fig. \ref{fig:RS_scaling}. Considering $R_r\approx 1$, simple
manipulations lead to
\begin{equation}\label{eq:group_size_scaling_approx}
  \left<\frac{n_{r-1}}{n_r}\right> \approx
  \frac{\<k_r\>}{\<k_{r+1}\>} = y^{s_r - s_{r+1}}.
\end{equation}
This result states that a constant group-size scaling in the outer
layers is possible only if the cost difference between them is
constant.
\begin{equation}\label{eq:equispaced_cost_condition}
  s_r-s_{r+1}=\Delta,\qquad (\text{for  } \,r\, \text{  s.t.  } \,n_r \gg 1).	
\end{equation}
%


\section{Fit to an empirical social network}
\label{ss:RS_dunbar_structure}
\begin{figure}[t]
  \centering
  \includegraphics[width=1.1\linewidth]{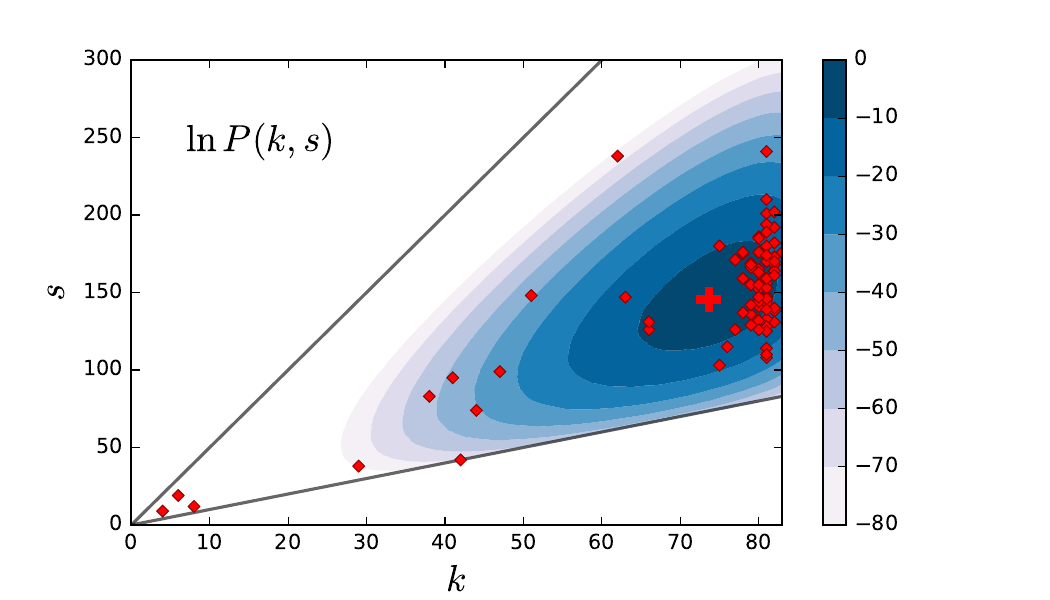}
  \caption{Contour plot for the joint log-density of states in
    the $k$-$s$ space for the grand-canonical ensemble with
    parameter values $\lambda=-0.3076$ and $\mu= -0.5842$.  The
    red diamonds correspond to the dataset individual
    ego-networks and the red cross marks the empirical
    averages. The lines $s=k$ and $s=5k$ delimit the
    configuration minimum and maximum allowed total weight,
    respectively.}
  \label{fig:DOS}
\end{figure}

\begin{figure}[b]
  \centering
  \includegraphics[width=0.7\linewidth]{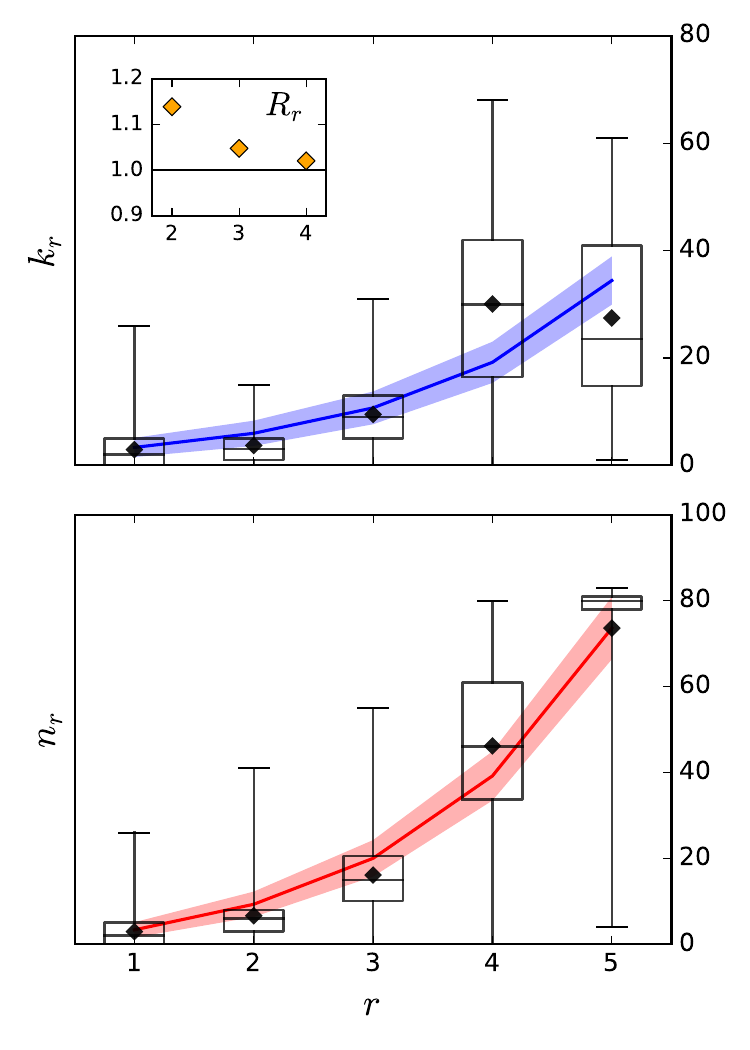}
  \caption{Layer degrees (top) and layer group sizes (bottom)
    distributions.  The empirical distributions are represented by the
    black box-plots, where the box comprises the second and third
    quartiles, separated by the median line. The whiskers extend to
    the full distribution domain, and the averages are represented by
    the black diamonds.  The colored lines join the corresponding
    ensemble averages, $\<k_r\>$ (blue, left), and $\<n_r\>$ (red,
    right) from the grand-canonical ensemble with parameters $N=84$,
    $x=0.74$ and $y=0.56$. Shaded regions comprise one standard
    deviation. The inset shows the numerical value of the
    micro-canonical correction factor
    $R_r=(1+\epsilon_r)/(1+\epsilon_{r+1})$ which tends to $1$ for
    the outer layers.}
  \label{fig:RS_scaling}
\end{figure}

The grand-canonical ensemble presented above may function as a null
model for ego-networks, providing a benchmark against which to test
more complicated features. In order to accept the ensemble as a good
model for social structure, the model should meet two demands:
\emph{(i)} It should generate ego-network instances with $k$ and $s$
values similar to the empirical ones (similar macrostate); and
\emph{(ii)} Those instances should present a nested layer
structure. The assumption of constant layer costs makes the ensemble
specially suited to model data from surveys, where the ties weights
are chosen from predefined discrete scores or categories.
We have fitted the grand-canonical ensemble to the Reciprocity Survey
(RS) dataset \cite{AlmaatouqShmueliPONE2016}. In this experiment, a
total of $N=84$ undergraduate students were asked to score their
relationship with each of the other participants in a scale from $0$
to $5$, where $0$ meant no-relationship, and $1$ to $5$ represented
increasing degree of friendship.
We have considered the zero weight as a no-link. Thus, the allowed
layer costs are $\{s_1,s_2,s_3,s_4,s_5\}=\{5,4,3,2,1\}$, which verify
the equispaced layer cost condition,
Eq. \eqref{eq:equispaced_cost_condition}.
The global and hierarchical structure of the empirical network is
summarized in Table \ref{table:RS_dunbar_structure}. The individual
ego-networks show the expected Dunbar's structure, albeit with some
remarks.
By design, the total active network is incomplete as the maximum
possible degree is limited by the total number of experiment
participants, $k_{\max}=N-1$. Consequently, the outer layer degree,
$\<k_5\>$, departs from the expected scaling.
However, all the participants belong to the same course and live in
the same campus, hence we can expect that a significant fraction of
their actual social network is captured by the experiment. Indeed, the
inner groups show an almost constant ratio of approximately $0.4$,
which is consistent with the ratio of $\sim 1/3$ reported by larger
scale studies \cite{ArnaboldiFB2012, ArnaboldiTW2013,
  DunbarPasarellaSN2015, MacCarronDunbarSN2016}.
The degree distribution is peaked close to the number of participants,
with low variance. The weights distribution, however, is more
spread-out.
\begin{table}[t]
  \begin{tabular}{|c|l||c|c|c|c|c|c|c|}
    \hline
    \multicolumn{2}{ |c ||}{global stats.} & layers & $s_r$ & $\<k_r\>$ & $\sigma_{k_r}$ & $\ff{\<k_{r-1}\>}{\<k_{r}\>}$ & $\<n_r\>$ & $\left<\ff{n_{r-1}}{n_{r}}\right>$ \\
    \hline
    $N$ & 84 & 			       $r=1$ & $5$ &  $2.93$  & $3.49$  &        & $2.93$  &     \\
    \hline
    $\<k\>$	& 73.63 &	       $r=2$ & $4$ &  $3.68$  & $3.30$  & $0.80$ & $6.61 $ & $0.39$ \\
    \hline
    $\sigma_k$ & 17.17 &   $r=3$ & $3$ &  $9.50$  & $6.04$  & $0.39$ & $16.11$ & $0.40$ \\		
    \hline
    $\<s\>$ &  145.45 &        $r=4$ & $2$ &  $30.07$ & $16.78$ & $0.32$ & $46.18$ & $0.37$ \\
    \hline
    $\sigma_s$ & 42.71 &   $r=5$ & $1$ &  $27.45$ & $15.81$ & $1.10$ & $73.63$ & $0.61$ \\
    \hline
  \end{tabular}
  \caption{Global statistics and layer structure of the Reciprocity Survey dataset network \cite{AlmaatouqShmueliPONE2016}.}
\label{table:RS_dunbar_structure}
\end{table}
We have fitted the grand-canonical ensemble to the observed data, by
substituting the mean values of $\<k\>=73.63$ and $\<s\>=145.45$,
along with $N=84$; into Eqs. \eqref{eq:saddle_k_canonical} and
\eqref{eq:saddle_s_canonical}.
Solving the saddle point equations numerically, we obtained the
parameter values $x=0.74$ and $y=0.56$.
The resulting distribution, along with the data are shown in Figure
\ref{fig:DOS}. We have employed a Wang-Landau algorithm
\cite{WangLandauPRL2001, FischerAltmannPRE2015} in order to compute
the joint density of states in the $k$-$s$ space (the macrostate
space).  This function is defined as $P(k,s) = \sum_{\{k_r\}}
P(\{k_r\})\delta(k-\sum_r k_r)\delta(s-\sum_r k_r s_r)$, where
$P(\{k_r\})$ is the grand-canonical probability distribution from
Eq. \eqref{eq:canonical_prob}. We have made available online a Python
implementation of the algorithm
\footnote{It is straightforward to generalize a Wang-Landau algorithm
  to compute a joint bivariate density of states in an integer valued
  configuration space. The code in python is available at
  \url{https://manu-jimenez.github.io/2017/04/19/Wang-Landau-for-joint-DOS.html}}.
As it can be seen on the figure, the presence of various outliers with
very low $k$ and $s$ displaces the averages from the bulk of the
distribution. Other than that, the probability distribution is a well
behaved unimodal function, hence fulfilling our first demand,
\emph{(i)}.
Next, let us compare the RS layer structure with the layer structure
generated by the grand-canonical ensemble. Figure \ref{fig:RS_scaling}
shows the empirical layer degree and layer group sizes distributions
along with the ensemble averages.
Observing the RS layer distributions we find that the empirical
averages of the three first layers lie within one standard deviation
of the ensemble expected values, both for degrees, $k_r$ , and group
sizes, $n_r$. On the other hand, the outermost layers degrees suffer
from finite size effects. The discrepancy between ensemble and data is
however corrected for the accumulated variables, $n_r$, which follow
closely the ensemble scaling trend. Thus, condition \emph{(ii)} is
verified as well.
Remarkably, the obtained value of $y=0.56$ is comparable with the
empirical group size scaling for the outer layers,
$\<n_4/n_5\>=0.61$. Indeed, the approximation from
Eq. \eqref{eq:group_size_scaling_approx} is better for the larger
outer layers, where the correction factor $R_r$ tends to $1$, as
shown on the inset of Fig. \ref{fig:RS_scaling}.
We have also fitted the ensemble thermodynamic limit to the empirical
data. Interestingly, despite the small size of the RS network, the
grand-canonical ensemble results are barely distinguishable from its
thermodynamic limit. Both ensembles average layer degrees and group
sizes are equivalent, and only the variances of the outer layer
degrees are slightly larger in the thermodynamic limit.

\section{Discussion}\label{s:discussion}

We have proposed a grand-canonical ensemble as a null model for the
hierarchical structure of social networks.
The ensemble generates the observed nested structure of increasingly
large layers of decreasing link weight. Moreover we show that, at
least for the outer layers, a constant group size scaling between
consecutive group pairings is possible only if the difference between
costs of consecutive layers is constant.
In the thermodynamic limit, that is, when the number of actors is
large, the layer degrees are uncorrelated Poisson variables, which are
related through the group size scaling, $y$.
Interestingly, a recent paper providing more evidence on Dunbar's
theory shows the good fit of Poisson distributions to the
layer-specific degree distributions \cite{kordsmeyer2017sizes}.
In the case of the dataset analyzed, after fitting the average values
of the social bandwidth or resource $\<s\>$ and degree $\<k\>$, we
find that typical samples of the ensemble are similar to the empirical
ego-networks.

Interestingly, the statistical properties of our null model bear a
strong formal relation to Bose-Einstein statistics. Such analogies
have been reported earlier in the literature, e.g. the
Bianconi-Barabási model for preferential attachment in networks
\cite{Bianconi}, where each node has a different energy level, and the
role of particles is played by the links, and presenting Bose-Einstein
condensation in a certain regime. Thus, it is relevant to ask whether
our model can reach this phase, in which a larger fraction of the
acquaintancies is accumulated at the lowest energy level, giving rise
to a social network dominated by {\em weak ties} \cite{Meo.14}. We
intend to consider this intriguing feature in future work.

The proposed null model succeeds at modeling survey social data, where
the available categories could be directly used as representations of
layers. However, in larger scale studies of ego-network social
structure, layers are not given \emph{a priori}, but rather
\emph{inferred} from the interaction patterns. The link weights
usually represent frequency of interaction which acts as a proxy for
emotional closeness, and ties are then clustered into discrete groups
according to these weights \cite{ArnaboldiFB2012, ArnaboldiTW2013,
  DunbarPasarellaSN2015, MacCarronDunbarSN2016}.
It is important to assert that we do not identify our link weights, or
costs, with interaction frequency. Rather, we introduce the abstract
notion of social resource, which can be spent in assigning discrete
weights to the social ties.
In order to justify this construction it can be argued that a limited
social bandwidth may arise both from biological or temporal
constraints, since maintaining a social relationship is costly
\cite{dunbar1992, GoncalvesVespignaniPONE2011}.
Nevertheless our model remains uninformative about the psychological
or sociological nature of that cost.
The other main assumption of our model is the discrete nature of the
layers. It is important to stress that we do not intend to neither
justify nor provide any sort of explanation to their existence. We
rather rely on the existing literature, where these layers have been
consistently identified and even given specific names: the
\emph{support clique} of size $\about5$, the \emph{sympathy group} of
size $\about15$, the \emph{affinity group} of size $\about50$, and the
total active network whose size equals the Dunbar's number,
$\about150$ \cite{ZhouDunbar2005}.
Assigning the layers a constant cost is not only a convenient
simplification, but has also a rather natural interpretation. Indeed
the very existence of the layers would implicitly define different
\emph{types} of relationships for the ego. We simply consider that
relationships within a given layer are similar precisely because they
have the same cost. Then, the problem we focused on was to compute in
how many different ways a given number of actors can be distributed in
an ego's network with the above mentioned restrictions.
In our setting, once the layers, and the total average degree and
social resource are fixed, the hierarchical structure emerges in a
natural way, as the number of possible configurations with few strong
and many weak links is much higher than configurations made of only
strong or weak links. This result suggests that the observed
hierarchical structure could be a consequence of the existence of an
internal discrete categorization in which individuals organize
different types of relationships. However, the reason why those
categories or layers would exist in the first place remains unknown
and shall be further explored.

\begin{acknowledgements}
  We acknowledge useful discussions with Ignacio Tamarit, Ignacio
  Morer, Conrad P\'erez-Vicente and Albert D\'iaz-Guilera. This work
  was supported by the Spanish Government through grants
  PGC2018-094763-B-I00 (SNS and EK), FIS2015-66020-C2-1-P (SNS) and
  FIS2015-69167-C2-1-P (JRL).
\end{acknowledgements}


\vspace{1cm}

\appendix

\section{Cumulants of a quotient function}\label{s:appendix_quotient_cumulants}
The cumulants of a quotient function $r(x,y)=x/y$ cannot be obtained
directly from the partition function. However, they can be
approximated by expanding $r(x,y)$ around the expected values, $\<x\>$
and $\<y\>$, as done in reference \cite{SagarraGuileraEPL2014}. The
mean and variance up to second order are given by
\begin{align}
\centering
 \<r\> &=\ff{\<x\>}{\<y\>}\[1 + \ff{\<y^2\>}{\<y\>^2} - \ff{\<xy\>}{\<x\>\<y\>} \],& \\
 \sigma^2_r &= \ff{\<x\>^2}{\<y\>^2}\[ \ff{\<x^2\>}{\<x\>^2} + \ff{\<y^2\>}{\<y\>^2} - \ff{2\<xy\>}{\<x\>\<y\>} \].&
\end{align}


\end{document}